\documentclass[journal,10pt]{IEEEtran}

\usepackage{amsmath, amssymb}
\usepackage{xcolor}
\usepackage{cite}
\usepackage{enumitem}
\usepackage{url}
\usepackage{bbm}
\usepackage{csquotes}
\usepackage{booktabs}
\usepackage{multirow}
\ifCLASSINFOpdf
   \usepackage[pdftex]{graphicx}
   \graphicspath{{Figures/}}
\else
\fi

\font\myfont=cmr10 at 21 pt

\makeatletter
\pretocmd\@bibitem{\color{black}\csname keycolor#1\endcsname}{}{\fail}
\newcommand\citecolor[1]{\@namedef{keycolor#1}{\color{black}}}
\makeatother

\begin{document}
\bstctlcite{IEEEexample:BSTcontrol}
\title{\myfont Indoor Positioning in 5G-Advanced: Challenges and Solution towards Centimeter-level Accuracy with Carrier Phase Enhancements}
\author{Jakub~Nikonowicz, Aamir~Mahmood,    Muhammad~Ikram~Ashraf, Emil~Björnson,
Mikael~Gidlund
\vspace{-24pt}
}

\maketitle
\thispagestyle{empty}

\begin{abstract}
After robust connectivity, precise positioning is evolving into an innovative component of 5G service offerings for industrial use-cases and verticals with challenging indoor radio environments. In this direction, the 3GPP Rel-16 standard has been a tipping point in specifying critical innovations, followed by enhancements in Rel-17 and Rel-18. In this article, we elaborate on the 5G positioning framework, measurements, and procedures before shifting the focus mainly to recently identified carrier-phase (CP) measurements in Rel-18 as a complementary measure for time- and angular-based positioning methods. We discuss the associated challenges and potential solutions for exploiting CP, including integer ambiguity, multipath sensitivity, and signaling aspects. Furthermore, we study how phase-continuous reference signaling can counter noisy phase measurements using realistic simulations to achieve centimeter-level accuracy in indoor factory (InF) scenarios.

\end{abstract}



\section{Introduction}
\IEEEPARstart{C}{ontinuous} work on the fifth-generation (5G) network aims to extend its applications beyond traditional mobile broadband, where one of the fundamental features is precise positioning. 
The need for high-precision positioning is anticipated in logistics, autonomous harbors and vehicles, localized sensing, digital twins, augmented and virtual reality, and more. It is desirable to deliver positioning services using cellular technology, instead of requiring a dedicated infrastructure. The 5G network design is evolving mainly in response to the needs of industrial segments, i.e., Industry~4.0. We can foresee the requirement for centimeter accuracy positioning in fully automated factories for acquiring precise knowledge of resources placement, track moving assets and machinery, up to product storage. As a result, smart factories remain the primary sector for precise positioning, and an indoor factory (InF) environment defines current requirements and challenges~\cite{Dwivedi2021}.

In this respect, the 5G new radio (NR) provides numerous innovations, from new positioning architecture and reference signals to measurement/parameter enhancements, for achieving positioning accuracy down to the centimeter. The 3GPP Rel-16 targets precise positioning and low-latency requirements of diverse application scenarios, including Industrial Internet-of-things (IIoT), transportation, and logistics. For instance, the indoor positioning accuracy (horizontal and vertical) and latency requirements for commercial use-cases are less than 3m, with less than 1s of latency and 80\% service availability. Meanwhile, the Rel-17 aims to improve the positioning accuracy up to tens of centimeters and lower latency up to tens of milliseconds for InF-IIoT.

The 5G positioning methods, in general, are derived from timing, angular, power-based techniques, and their combinations wherein Rel-17 proposes improved signaling and procedures over Rel-16 considering that i) wider bandwidth increases timing measurements resolution, ii) larger antenna array apertures in massive MIMO (multiple-input multiple-output) for mid-band and mmWave allows narrower radio beams and increased angular resolution, and iii) joint processing of time-and angular-based methods helps mitigate dense multipath in InF settings.

\begin{figure*}[!t]
    \centering\includegraphics[width=0.95\linewidth]{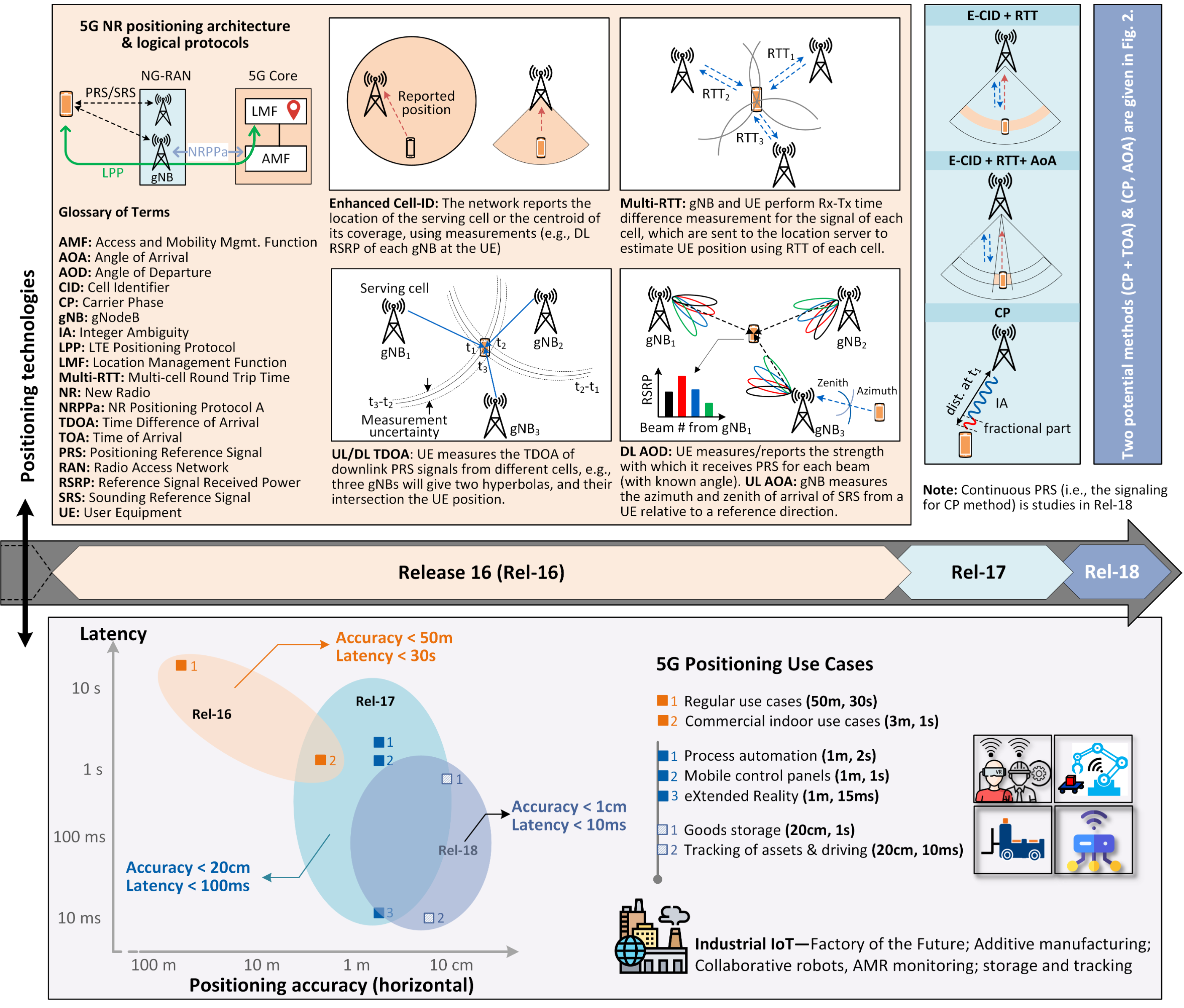}
    \vspace{-8pt}
    \caption{The 5G positioning architecture and possible measurements/methods together with performance requirements across 3GPP releases for evolving needs of IIoT use-cases. Rel-17 and onward efforts are focused on accuracy enhancements by improving NR parameters (e.g., bandwidth, power, antennas), combination of measurements (e.g., CP with time/angle measurements), and other countermeasures (e.g., multipath resolution).}
    \label{fig:MainFig}
\end{figure*}

The current focus of the ongoing Rel-18 study item~\cite{RP-222616} is to investigate solutions to enhance accuracy up to a few centimeters within a few milliseconds of latency. This involves i) bandwidth aggregation for intra-band carriers, ii) integrating sidelink information, and iii) identifying reference signals, physical layer measurements/procedures to enable carrier phase (CP)-based positioning. Fig.~\ref{fig:MainFig} shows the 5G positioning methods, as well as the evolving requirements and emerging use-cases in 3GPP releases~\cite{38.857}.

The earlier studies (e.g., \cite{Dwivedi2021, 38.857}) are focused on general description of the 5G positioning architecture and methods proposed in Rel-17\&18. Meanwhile, the study \cite{fouda2022toward} analyzed the CP-based positioning, which is finding the traction to complement time/angle-based methods and eliminate their sensitivity to sampling resolution and system parameters.

However, the various entailing critical issues of realizing CP-based positioning enhancements are not considered in these works. In this respect, our main contributions are:
\begin{itemize}[leftmargin=*]
    \item To establish the relationship between NR system parameters and delay/angle error variance, we describe the 5G positioning architecture, including reference signals and basic positioning measurements/methods.
    \item We introduce CP-based enhancements in time/angle-based positioning schemes along with related problems and challenges.
    \item We consider CP measurement enhancements under continuous reference signals as a potential solution to counter noisy phase measurements.
    \item We provide simulation results for CP measurements, with/without continuous reference signals, in estimating 3D distance error in the InF channel models.
    \item Finally, further improvements/benefits in continuous CP measurements are envisioned.
\end{itemize}

\section{Basic positioning approaches}
The 3GPP specifications entail the basic architecture, main positioning technologies, and possible combinations (see Fig. ~\ref{fig:MainFig}), which are briefly resumed here.

{
\renewcommand{\arraystretch}{1.1}
\begin{table*}[ht]
\centering
\caption{Relationship between 5G NR parameters to delay/angle error variance~\cite{Dwivedi2021, RD_TOA_Est_TWC}.}
\label{tab:EstimatesRelations}
\scalebox{0.9}{
\begin{tabular}{@{}llll@{}}
\toprule
\textbf{Estimate}  & \textbf{Parameter} & \textbf{Relationship} & \textbf{Comments}\\ 
\toprule
\multirow{8}{*}{\textbf{Delay error variance}} & Bandwidth (BW) & $\propto{1}/{\text{BW}^2}$ & \begin{tabular}[c]{@{}l@{}}NR typically provides 100\,MHz bandwidth at FR1 and \\ 400\,MHz at FR2\end{tabular} \\
\cline{2-4}
 & Average received power (P) & $\propto {1}/{\text{P}}$ & Received power is also increased by the array gain\\
 \cline{2-4}
& Subcarrier spacing (SCS) & $\propto {1}/{\text{SCS}}$ & \begin{tabular}[c]{@{}l@{}}Higher SCS increases resolution in resolving LOS from\\the multipath\end{tabular}\\
\cline{2-4}
& \multicolumn{3}{l}{\textbf{Antenna array geometry}}\\
& \begin{tabular}[c]{@{}l@{}} Antenna elements (N) 
\end{tabular} & \begin{tabular}[c]{@{}l@{}} $\propto {1}/{\text{N}}$ 
\end{tabular} & Only the array gain matters, so the array geometry can be arbitrary\\ 
\midrule
\multirow{8}{*}{\textbf{Angle error variance}} & Bandwidth (BW) & -- & Independent\\ 
\cline{2-4}
& Average received power (P) & $\propto {1}/{\text{P}}$ & Less pronounced compared to delay error variance \\
\cline{2-4}
& Subcarrier spacing (SCS) & $\propto {\text{SCS}}$ & \begin{tabular}[c]{@{}l@{}}For fixed BW, increasing SCS leads to fewer details in \\the channel frequency response\end{tabular}\\
\cline{2-4}
& \multicolumn{3}{l}{\textbf{Antenna array geometry}}\\
& \begin{tabular}[c]{@{}l@{}} Horizontal antenna elements ($\text{N}_\text{H}$)\\ Vertical antenna elements ($\text{N}_\text{V}$)\\ Antenna spacing (S)\\ 
\end{tabular}    & \begin{tabular}[c]{@{}l@{}} $\propto {1}/{\text{N}_\text{H}^3}$ \\ $\propto {1}/{\text{N}_\text{V}^3}$ \\$\propto {1}/{\text{S}^2}$ 
\end{tabular} & \begin{tabular}[c]{@{}l@{}} The array geometry matters, the number of horizontal and vertical\\ elements affect differently depending on the location of the device \end{tabular} \\
\bottomrule
\end{tabular}}
\vspace{-10pt}
\end{table*}}

In the 5G positioning framework, the location management function (LMF) is a central entity to estimate UE position based on assistance/measurements from next-generation (NG)-RAN and UE through the access and mobility management function (AMF). Moreover, a new NR positioning protocol A (NRPPa) carries the positioning information between NG-RAN and LMF. 
Besides, NR introduced two reference signals specifically for enabling accurate positioning measurements in downlink (DL) and uplink (UL) directions, respectively, as positioning reference signal (PRS) and sounding reference signal (SRS). These reference signals are configurable to enhance the precision of UL/DL time/angular measurements (see Sec.~\ref{subsec:cprs}). LMF provides DL-PRS configuration to UEs using LTE positioning protocol (LPP) while RAN configures UL-SRS to UEs using radio resource control (RRC) protocol. To understand the relations of these measurements with the system parameters, Table~\ref{tab:EstimatesRelations} provides a summary of the various factors affecting errors in time/angle-based positioning.

\subsection{Angle-based Positioning}
\label{subsec:angle_pos}
Massive MIMO and smart antenna techniques in 5G allow precise control over the beamforming and angle estimation. Angle-of-departure (AOD) and angle-of-arrival (AOA) measurements are part of the beam management procedure specified by 3GPP. Based on UL or DL transmission direction, two positioning techniques are:
\begin{itemize}
    \item \textbf{DL-AOD:} using PRS, a UE measures the beam reference signal received power (RSRP) of gNB and reports it to LMF via LPP~\cite{Dwivedi2021}. 
    \item \textbf{UL-AOA:} using SRS, gNB measures AOA, i.e., the azimuth and elevation angles, and reports them to the LMF by the NRPPa~\cite{Dwivedi2021}.
\end{itemize}

The achievable accuracy of triangulation based on angle measurements is limited due to the range/resolution of reportable absolute values for power measurements, which is $[-156, -31]$dBm, with 1dB resolution \cite{Dwivedi2021}. Moreover, RSRP-based approaches require establishing an accurate propagation model to reliably estimate signal energy, challenging high measurement accuracy in dynamic radio environments and limiting the scope of AOA/AOD-based techniques~\cite{Zhang2021}. Therefore, there are many suggestions to improve AOA/AOD performance; for instance, by introducing CP measurement in addition to beam RSRP measurement \cite{R1-2104844} (see Sec.~\ref{subsec:cpa}).

\subsection{Time-based Positioning}
\label{subsec:time_pos}
An alternative solution is trilateration methods. Time-of-arrival (TOA)-based positioning transforms signals propagation delay into the distance between gNB and UE. However, this approach requires precise UE-gNB synchronization. TDOA is an improved ranging method based on differential TOA, requiring time synchronization between gNBs only. As with the angular methods, the PRS and SRS are used to estimate the DL or UL TDOA between different gNBs. 

\begin{itemize}
    \item \textbf{DL time-difference-of-arrival (DL-TDOA):} UE receives the PRS from several gNBs and calculates the TOA of each PRS signal. The TOA of one gNB is taken as a reference to compute the reference-signal-time-difference (RSTD) to TOAs from the remaining gNBs. UE sends the RSTD measurements to LMF to compute the UE position using known geographical coordinates of gNBs. 
    \item \textbf{UL time-difference-of-arrival (UL-TDOA):} UE transmitted SRS is received by neighboring gNBs; a transmission measurement function calculates the relative-time-of-arrival (RTOA) and sends it to the LMF to compute the UE position.
    \item \textbf{Multi-cell round-trip-time (Multi-RTT):} gNB and UE perform Rx-Tx time difference measurement, using PRS and SRS signaling, for the signal of each cell. LMF initiates the procedure whereby multiple gNBs and a UE perform the gNB Rx-Tx and UE Rx-Tx measurements, respectively.
    Multi-RTT has higher accuracy than TDOA-based methods and relaxes requirements on time synchronization \cite{Dwivedi2021}.
\end{itemize}

Clock synchronization between gNB and UE or between gNBs is required to accurately estimate TOA or TDOA, respectively. However, the precision of time measurements is limited to intervals of $T_c$, with a flexible resolution of $2^{k}T_c$, where $T_c=0.51$ns and $k$ is an integer in the interval [2, 5] for FR1 and [0, 5] for FR2~\cite{Dwivedi2021}. Additionally, a dense multipath environment can cause non-line-of-sight (NLOS) propagation, making the measurements unreliable. Thus, the measurement quality critically depends on system timing and multipath propagation conditions~\cite{Rappaport2021,Fan2022}.

\subsection{Investigated Hybrid Solutions/Enablers}
The measurement types discussed earlier offer several possibilities for developing hybrid positioning solutions. For instance, the authors in \cite{mogyorosi2022} have summarized state-of-the-art approaches, including fingerprinting-based localization, which involves creating a database by measuring signal or antenna attributes at known locations and using machine learning models to estimate the position based on attributes measured by the mobile unit. An alternative to this method, which uses noisy AoA and ToA measurement models, is to employ an extended Kalman filter positioning engine or a distribution-based unscented Kalman filter. Another approach involves creating and reusing channel models based on real 5G measurements to enhance power-delay-profile-based algorithms. Machine-learning-aided positioning solutions constitute a separate group of methods being investigated. A supplementary strategy is assisted positioning that leverages 5G signals aided by GNSS or visible light communication.

Apart from mixed solutions, a 3GPP study indicated two other methods considered in Rel-18 for enhanced accuracy (see 3GPP TR~38.859): 1) PRS/SRS bandwidth aggregation for intra-band contiguous carriers feasible for single-chain Tx/Rx architectures, 2) NR carrier phase positioning with the potential to achieve horizontal accuracy of a few centimeters using existing PRS and SRS signals under certain conditions. 

\section{Carrier Phase (CP)-based Enhancements}

\begin{figure}[!t]
    \centering\includegraphics[width=0.95\linewidth]{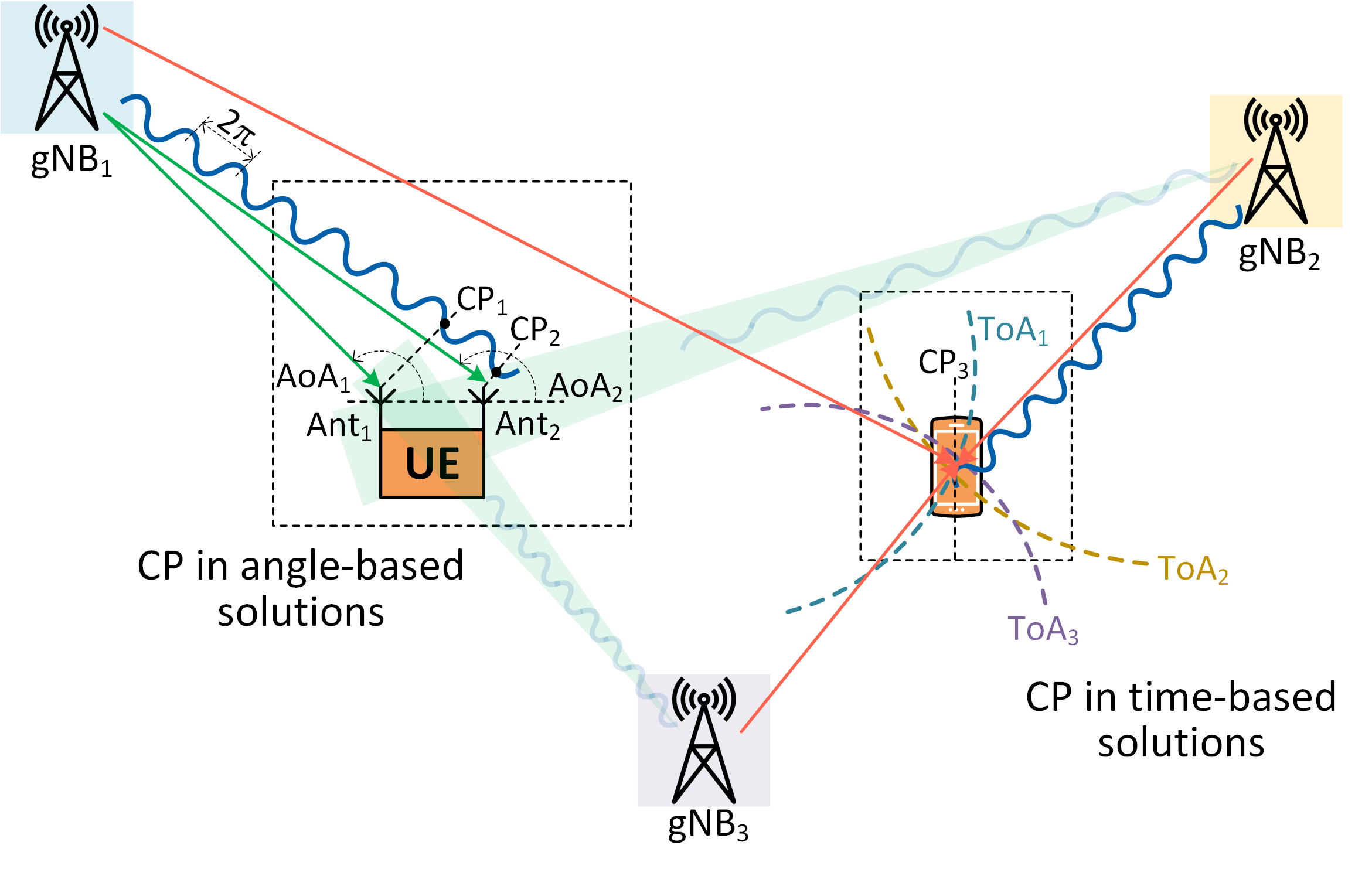}
    \caption{Illustration of carrier phase (CP) in angle/time-based solutions.}
    \label{fig:CPinOthers}
    \vspace{-14pt}
\end{figure}

Carrier phase-based positioning enhancement provides another measurement type for developing hybrid solutions. Generally, the CP measurement captures the difference between the phase of the incoming carrier signal and the receiver-generated reference signal. Under line-of-sight (LOS) conditions, the phase is in the $[0, 2\pi)$ range, and the measurement error is only a small fraction of the wavelength; reaching the centimeter level. As illustrated in Fig.~\ref{fig:CPinOthers}, CP-measurements can improve the accuracy of trilateration-based (i.e., using solely timing measurements) and triangulation-based (i.e., using AOD/AOA measurements in multi-antenna systems) positioning~\cite{R1-2104844}.

In positioning within InF settings, NR tackles challenges akin to those encountered in GNSS applications. Therefore, background from GNSS solutions can offer valuable insights, e.g., pseudolite/repeater positioning systems, transmitting GNSS-like signals, and strategically amplifying coverage where direct GNSS is unavailable (for an in-depth review of GNSS-based indoor solutions, refer to \cite{li2023}). Indoor satellite positioning operates with phase differential measurements on clean (demodulated, data-whipped, and continuous) L1 and L2 carrier signals for accurate positioning to sub-wavelength levels of about 20\,cm).
Wireless networks offer advantages over GNSS due to adaptable carrier frequencies and error mitigation, creating an opportunity for mm-level accuracy. Furthermore, 5G networks provide a platform for various services beyond positioning; hence, positioning as a service exploits shared infrastructure, potentially reducing costs, unlike GPS pseudolite stations which require separate setup and maintenance. Yet, applying carrier-phase measurements to wireless networks requires pre-adaptation due to intricate features like intermittent signal blocks and cyclic prefixes within OFDM signals.

\subsection{CP in Time-based Solutions}
As mentioned earlier, PRS are pseudo-random sequences with good autocorrelation. The UE correlates the time domain samples with the known PRS pattern when measuring the signal propagation delay. After detecting the code phase, the receiver replicates the transmitted pseudo-random code and moves the replica until the maximum correlation is achieved. The offset corresponds to the transmitter-to-receiver signal propagation time. Hence, it depends on estimating the earliest peak delay in the magnitude of the normalized cross-correlation function. Consequently, the accuracy of the range-based methods depends mainly on the UE sampling resolution and the signal bandwidth~\cite{Zhang2021}.

Contrary to delay in code-phase detection, carrier-phase detection translates the phase difference and wavelength into a distance. When the receiver intercepts the transmitted signal, it is locked by a phase-locked loop (PLL). From the moment the first path is received, the phase shift between the locally generated reference signal and the replica of the received carrier wave is constant. Therefore, the measuring point is not as critical to the measurement quality~\cite{Fan2022}. However, when initiating positioning, the received phase is in the range $[0, 2\pi)$, so only a fraction of a single wavelength can be measured. This causes integer ambiguity (IA) problem---how many full wavelengths ($N$) precede the measured fraction at the propagated distance~\cite{Zhang2021} (see Section \ref{subsec:ia}).

\subsection{CP in Angle-based Solutions} 
\label{subsec:cpa}
Carrier-phase (CP) measurement can also improve positioning accuracy by triangulation. Accurate measurement of an angle in a plane requires a configuration with (at least) two receiving antennas. The transmitter emits a sinusoidal signal with wavelength $\lambda$ that propagates spherically so that the wave phase changes continuously in $[0, 2\pi)$. The instantaneous signal phases $\phi_{1}(t),\phi_{2}(t)$ measured at the two antennas will generally be different, but the difference $\Delta=\phi_{1}(t)-\phi_{2}(t)$ is constant and enables AOA estimation.
Suppose the separation between the receive antennas is $d$ and that the impinging wave is planar. The AOA $\theta$ perpendicular to the line between the antennas satisfies $\cos(\theta) = \Delta \lambda/(2\pi d)$, from which the angle can be extracted (but there are multiple solutions).
Similar techniques can be used to measure the AOD from a multi-antenna transmitter to a single-antenna receiver; for example, the transmitter antennas can take turn in sending the sinusoid so that the receiver can measure the respective phase-shifts.

Moreover, the CP method is simpler and more efficient than the beam-sweep method. It provides an accurate angle estimate, and simultaneously, multiple terminals can acquire AOA/AOD information from a single PRS transmission. Precise angle measurement using a dual-antenna configuration is also a solution proposed in Bluetooth-SIG~\cite{R1-2104844}.

\subsection{Continuous-PRS}
\label{subsec:cprs}
The implementation of the CP-based positioning enhancement is based on the PRS signal transmitted by gNB. 
PRS is designed to support DL-based positioning schemes while delivering accuracy, coverage, and interference management. PRS can cover the full NR bandwidth, and its transmission over multiple symbols (using comb-size of 2, 4, 6, and 12, i.e., the density of subcarriers in a PRS symbol) allows accumulating power. It can start at any physical resource block (PRB), configured in steps of 4-PRBs from 24 to 276 PRBs, giving a maximum bandwidth of 100\,MHz for 30\,kHz subcarrier spacing (SCS) in FR1 and 400\,MHz for 120\,kHz SCS in FR2~\cite{Zhang2021, Dwivedi2021}.

The current PRS structure enables the efficient determination of TOA and phase-of-arrival (POA). While a single measurement indicating a code-based shift is essential in TOA, in POA, a phase shift remains constant and measurable for the signal duration and thus can be analyzed/tracked continuously for phase noise reduction. Therefore, a continuous PRS (C-PRS) structure is advantageous compared to the current shifted PRS configuration pattern.
Regardless of where the FFT sampling window starts during the signal duration, the C-PRS signals maintain orthogonality, reducing noise in phase comparison. Moreover, when PRS signal sets from neighboring gNBs arrive with different delays, signals from distant gNBs may interfere, and only a part of the signal may be included in the FFT window. By providing C-PRS, the receiving UE may discard the boundary symbols, including the error-prone part, and collect only the intermediate symbols containing the full wavelength of the subcarriers. It is also essential that multiple oversampling by shifting the FFT window over the entire signal interval gives an identical result. This carries various implications:
\begin{itemize}
    \item The relative phase difference between the subcarriers is constant; thus, the phase difference between the PRS signals of different gNBs remain constant regardless of the position of the sampling window.
    \item Constant difference between the gNBs signals allows optimizing the differential measurement that solves the UE phase mismatch problem. 
    \item Iterative sampling results can be averaged to reduce noise, improve gain, and increase angular resolution.
\end{itemize}

\begin{figure}[!t]
    \centering\includegraphics[width=0.95\linewidth]{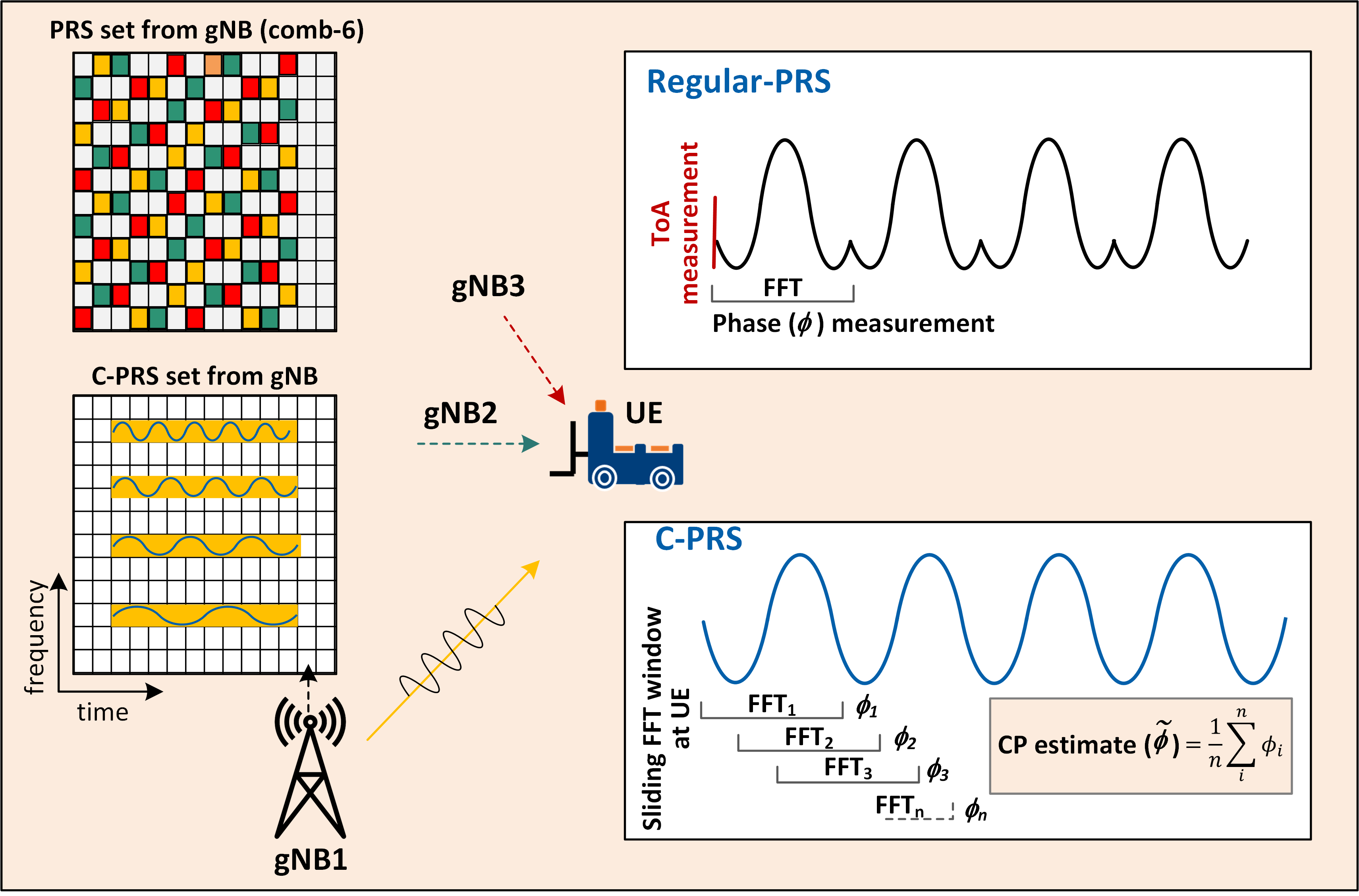}
    \caption{Carrier phase (CP) measurement in regular PRS vs. continuous-PRS.}
    \label{fig:CPRS}
    \vspace{-10pt}
\end{figure}

With today's rapid network development, it is becoming increasingly important to reuse existing hardware and possibly signals for both detection and communication functions. However, the current PRS structure, with cyclic prefix and scattering of symbols in PRBs, makes it impossible to benefit from the advantage of the signal continuity. Due to the cyclic prefix, the end part of the symbol is copied and used as a prefix. As a result, most subcarriers are discontinuous at the symbol boundaries, hindering building a continuous waveform over multiple symbols.
To overcome this, 3GPP initiates discussion over C-PRS, either as a pure carrier wave of periodic wide-band sinusoidal signals or a continuous narrow-band signal transmitted at a pre-defined carrier frequency \cite{fouda2022toward}. To this end, work item \cite{R1-2104880} considers a block-type PRS with a modified prefix. It proposes a low complex in-band method where the desired tone signal can be easily generated in a sequence that rotates the subsequent symbol phases by the length of the cyclic prefix interval. This process enables a seamless connection of the subcarrier waveform at the symbol boundaries (see Fig.~\ref{fig:CPRS}).

Note that the phase tracking reference symbol (PTRS), designed to correct the clock's phase noise, also adopts a similar block-type symbol configuration pattern.

\section{Performance of Continuous-PRS in InF}
We designed an example simulation scenario to demonstrate the NR positioning enhancement achievable by CP measurements, including continuous carrier phase (CCP) measures. The simulation parameters follow TR~38.857 as the typical InF scenario applicable in the performance analysis of current positioning solutions:
\begin{itemize}
    \item \textbf{FR1-specific values:} carrier frequency 3.8\,GHz, bandwidth 100\,MHz, SCS 30kHz.
    \item \textbf{FR2-specific values:} carrier frequency 28\,GHz, bandwidth 400\,MHz, SCS 120\,kHz.
\end{itemize}
The simulation scenarios FR1 and FR2 shared common values, i.e., PRS sequence generated for 3276 subcarriers with a comb-6 pattern, complex Gaussian noise with power adjusted to 10dB SNR at the UE-Rx. The gNB and UE positions are fixed at (x, y, height) in meters as (100, 100, 15) and (120, 100, 1.5), respectively. Simulations demonstrate the prospect of standalone methods by specifying the distance error in measurement to a single anchor in three-dimensional space.

To evaluate the performance of communication systems during standardization, it is crucial to use a geometry-based stochastic channel model suitable for the propagation environment, i.e., specific close-in free space reference distance models with frequency-dependent path loss exponents. 3GPP TR 38.901 Indoor Factory model considers the influence of industrial production on channel impulse response and corresponding power delay profile. Simulated channel characteristics follow different versions of the InF conditions defined in 3GPP 38.901, i.e., LOS and NLOS with sparse (S) or dense (D) clutter.

TOA-based distance measurements are made directly from the time lag at the correlation peak. They are mainly limited by the sampling frequency, period corresponding to the chirp signal, and the presence of LOS. The comparison of LOS conditions for FR1 and 2 indicates a reduction of the variance in the TOA by broadening the band, as pointed out in Table \ref{tab:EstimatesRelations}. The simulated NLOS scenarios clearly show how seriously the distance error increases in the absence of a direct path. Moreover, apart from a fundamental phase error after wave reflections, NLOS erroneously extends the search space of IA, for which TOA creates an incorrect entry point. Therefore, CP-based enhancements should be limited to LOS conditions.
\begin{figure}[!t]
    \centering\includegraphics[trim={0 0 0 0.8cm},clip, width=1\linewidth]{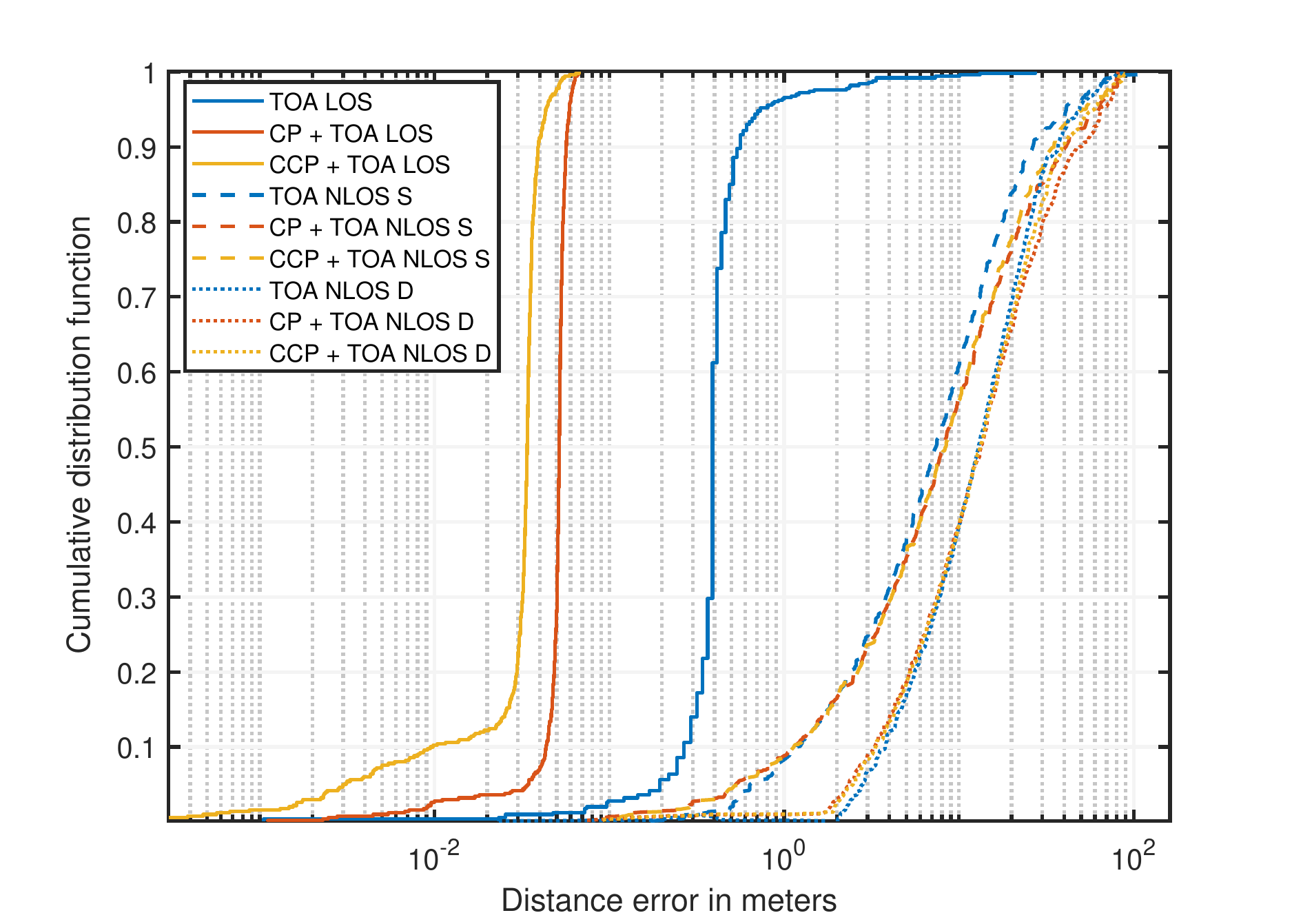}
    \caption{Cumulative distribution function (CDF) of the 3D distance error for the InF FR1 scenario with 100\,MHz bandwidth.}
    \label{fig:FR1}
\end{figure}

\begin{figure}[!t]
    \centering\includegraphics[trim={0 0 0 0.5cm},clip,width=1\linewidth]{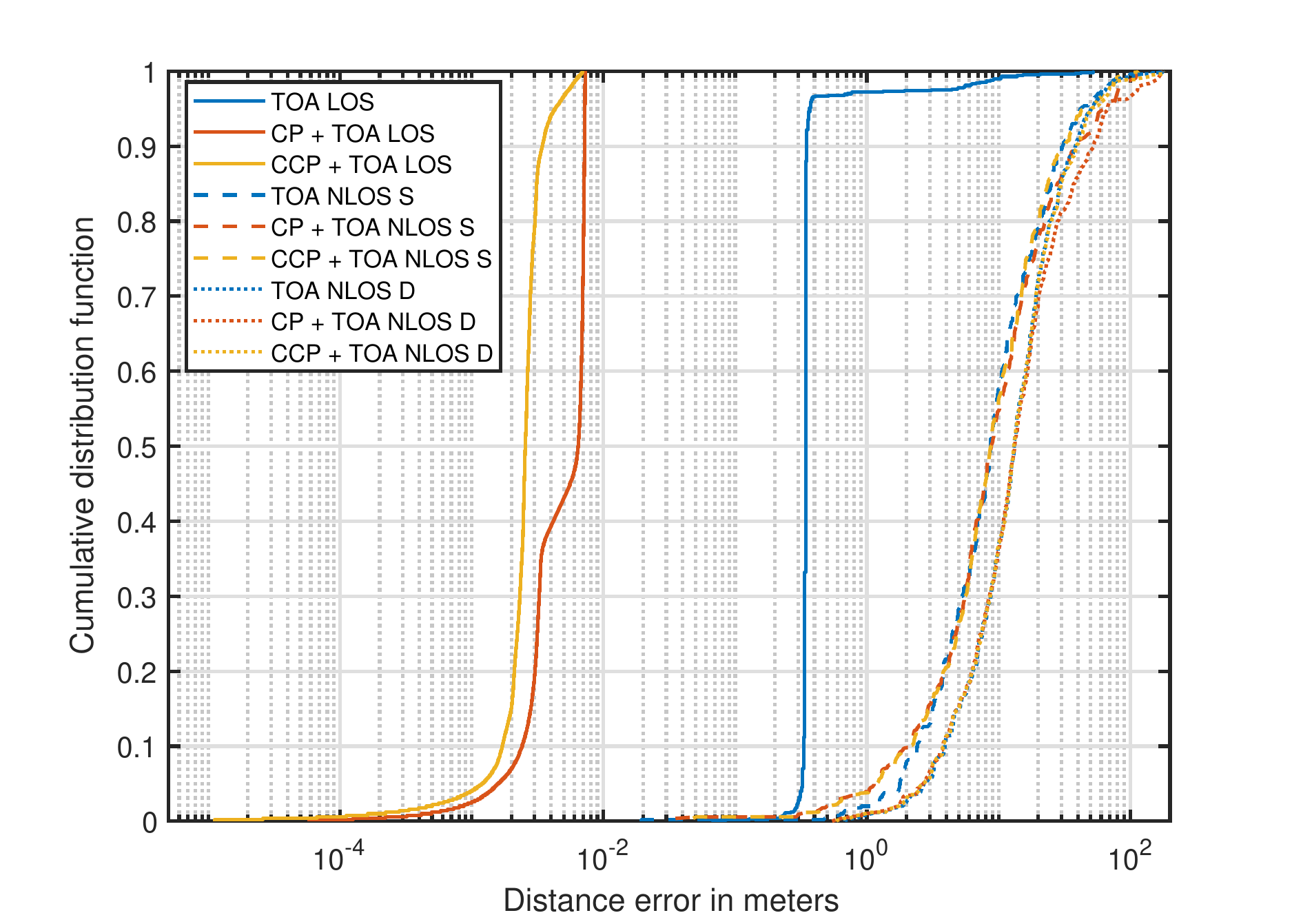}
    \vspace{-8pt}
    \caption{Cumulative distribution function (CDF) of the 3D distance error for the InF FR2 scenario with 400\,MHz bandwidth.}
    \label{fig:FR2}
    \vspace{-12pt}
\end{figure}

CP-based distance measurements follow the procedure of probing the phase of the middle subcarrier \cite{Zhang2021}, as proven to reflect reliable UE--gNB distance information. The sampling process also limits single-phase measurement, i.e., the actual information is rounded to the nearest sampling point. However, the phase noise inherent in propagation significantly influences distance accuracy within the range of a single subcarrier period.
Multiple phase measurements can provide phase noise reduction. CCP measurement assumes a constant phase difference over the received signal and the reference produced at the receiver.
Up to this point, simulation provides double replication and concatenation of the current OFDM symbol. Then, a 1000-fold FFT window sweep is used in the phase comparisons, with a 1-sample shift between measurements. In Figs.~\ref{fig:FR1} \& \ref{fig:FR2}, the performance of the continuous-PRS is reflected using the CDFs obtained for CCP, which shows the reduction of phase noise variance. 

An aspect that needs to be considered in phase folding is the possible phase shift resulting from the receiver mobility. In an indoor positioning system, a mobile receiver is moving towards a fixed transmitter with a typical speed of 3--9 km/h (0.83--2.5m/s), resulting in the frequency shift due to the Doppler effect will be 0.003--0.008ppm. Therefore, the resulting phase shift can easily be neglected for short, time-coherent periods.

\section{Open Challenges and Potential Solutions}
The ongoing work on positioning in 5G-and-beyond networks is dealing with a small set of problems that are entirely new. CP measurements, already used in outdoor systems, face challenges when migrating to InF environments due to cluttered propagation conditions, shorter distances, higher accuracy and latency targets, and a different signal structure. A significant part requires reevaluation of problems/solutions from some related fields.

\subsection{Integer Ambiguity}
\label{subsec:ia}
Introducing CP measurements in NR can reduce the measurement error, typically 10\% of the carrier wavelength under the right conditions, and consequently, increase the positioning accuracy. However, using CP is coupled with a difficulty, commonly referred to as integer ambiguity (IA); that is, the phase measurement is in the $2\pi$ range, but the total distance contains unknown integer numbers of the carrier wavelength. Consequently, a fast/reliable IA solution is essential to meet the low-latency positioning demands. Notably, the challenge posed by the unknown wavelength multiplicity between the transmitter and receiver has already been effectively addressed within GNSS CP measurements, laying the groundwork for adaptation in NR.

The main issue in IA is the large search range of possible integers, especially for the short wavelengths in the FR2 band. Using TOA can significantly reduce the IA search range or, alternately, time needed to solve it. Another approach to reduce the IA search range is the idea of virtual wavelength. Specifically, instead of transmitting a single frequency, the transmitter transmits reference signals in two or more frequencies to get phase measurements at multiple frequencies. As the two-phase equations follow the same propagation pattern, and only differ with wavelengths, represented by $\lambda_{1}$ and $\lambda_{2}$, both equations can be unified by using two-sided multiplication by $\lambda_{2}/(\lambda_{2}-\lambda_{1})$ and $\lambda_{1}/(\lambda_{2}-\lambda_{1})$, respectively. Then, subtracting the phase equations creates the virtual phase measurement for wavelength, $\lambda_{2}-\lambda_{1}$. This provides the opportunity to make differential wavelength much longer than the initial $\lambda_{1}$ and $\lambda_{2}$ \cite{fouda2022toward}. Adapting this solution in NR gives numerous possibilities as the network can configure the transmission frequencies to reduce the IA search space and computational overhead optimally. Therefore, further investigation into effective adaptation is required, and reevaluation of the sensitivity of estimation accuracy to IA errors caused by complex indoor propagation environments.

\subsection{Multipath and NLOS}
The IIoT use-cases exhibit harsh channel conditions caused by multiple reflections from objects (called ``clutter'') in a dense industrial environment. Thereby, the measurements suffer from a) disturbed phase continuity, b) excessive delays with respect to the LOS transmission time, and c) the angular deviation from the actual LOS direction. NLOS causes fundamental error in phase after reflection; simultaneously, a significant error is introduced into the location equations, and the IA search space vastly extends. Consequently, the centimeter-level accuracy in CP-based positioning becomes unattainable in desired time unless NLOS measurements are excluded or mitigated.

In LOS scenarios with multipath, the Rician-K factor measures the ratio of the received power from the LOS path to the received power from all other paths \cite{Rappaport2021}.
It is mainly the angle of the LOS path that is useful for positioning and antenna arrays can be used to distinguish it from the clutter. This is a classical signal processing problem that has resulted in the multiple signal classification algorithm (MUSIC), signal parameters via rotational invariant techniques (ESPRIT), and space-alternating generalized expectation maximization (SAGE) algorithms \cite{Gentile2013}. The associated computational complexity and robustness are challenging for real-time positioning. 

It can be challenging to distinguish a LOS scenario from a NLOS environment with one or multiple strong reflected paths. An estimated Rician-K factor can be clearly above 0dB but the dominant path might be a reflection from a misleading direction. Channel features such as maximum received power, delay spread, and departure/arrival angle spread can be used to distinguish between LOS/NLOS situations. An RSRP-based NLOS link identification algorithm is developed in~\cite{R1-2104909}, which performs binary hypothesis testing between LOS/NLOS link types using average NLOS channel power of the subcarrier.

Cooperative positioning is another approach, where erroneous distance measurements due to NLOS conditions are detected/corrected using information from neighboring anchors. However, a standalone device must be capable of autonomously detecting the NLOS state and correcting erroneous measurements only with local information. To this end, the concept of simple classifiers based on the received signal strength and first path information can be replaced by more advanced machine learning (ML) techniques~\cite{Rappaport2021}. ML-based solutions aggregate a variety of features. For instance, under NLOS conditions, multiple reflections cause the signals to be more attenuated and have lower energy and amplitude. While for LOS, the strongest signal path corresponds to the first arrival, for NLOS, the strongest path is preceded by weak components.

\subsection{UE and gNB timing errors}
Proper CP difference requires UE--gNB synchronization, analogous to TOA measurements. Herein, the differentiation of measurements known from TDOA is useful. Differential CP measurements allow to cancel-out UE clock offset, and eliminate typical measurement errors caused by the propagation environment. The differential measurement, therefore, assumes that gNBs are synchronized. However, the gNB subsets arrive at the UE with a slight time difference, and the phase measurement may be distorted/incorrect due to the gNBs' clock mismatch.
To support CP, all timing errors must be eliminated through double differencing measurements from two or more receivers/transmitters. Double differencing can mitigate the impact of clock frequency offset, oscillator drift, and initial phase errors. However, the target UE requires at least one positioning reference unit (PRU), e.g., one reference gNB, to measure signals from gNBs. The use of PRUs to facilitate CP will be evaluated in Rel-18 SI.

In respect to timing, C-PRS also provides prospects for further research through adaptation, including:

\subsubsection{\textbf{gNB/TRPs master-slave synchronization}}
Like the code phase-detection for TOA, the CP-detection translates the corresponding POA into a distance. Once the receiver intercepts the transmitted signal, it is locked in the PLL and allows the phase difference to be captured. However, the UE's local oscillator phase offset can significantly influence the measurement result. This forces, e.g., double differentiation in CP. Yet, for differential measurements, gNBs must be precisely synchronized (less than one carrier wavelength), which requires a sub-nanosecond variance.

It is difficult to synchronize gNBs to an ideal distant clock (GNSS) in sub-nanoseconds accuracy due to atmospheric fluctuations, temperature fluctuations, the inherent error of the device, etc., and positioning accuracy of less than 1m cannot be guaranteed~\cite{R1-2104880}. However, it is relatively simple for gNBs to follow a certain near external master clock in sub-nanoseconds variance. For this purpose, the discussed C-PRS can be used to trace the master clock phase continuously.

\subsubsection{\textbf{Digital femtosecond time difference}}
Improving the phase measurement resolution already has promising solutions in closely-related areas, such as the synchronization of complex laboratory infrastructure. The digital dual mixer time difference (D-DMTD) is a digital femtosecond time difference circuit developed in CERN~\cite{Sandia2017}. D-DMTD measures the phase difference between two digital signals with very accurate resolution using a relatively low-frequency counter. 
By sampling a signal of a particular frequency with a slightly slower clock, D-DMTD stretches the signal in the time domain, allowing intensive aliasing of the two input clocks fed to the phase detector with femtosecond time resolution. 

\subsection{Near-field positioning}
When an antenna array becomes sufficiently large, compared to the wavelength, the spherical curvature of the impinging wave becomes noticeable. Since the curvature depends on the propagation distance, this feature enables ranging in addition to conventional AOA estimation, so a single array can localize the transmitter. This operational regime is called the radiative near-field and exists for ranges up to the Fraunhofer distance. The combination of physically large arrays and the use of mmWave bands can jointly extend the near-field to ranges larger than a kilometer. A key implementation challenge is to keep a sufficient phase-synchronization across the array to enable accurate estimation of the curvature.

\section{Conclusions}
This article provides an overview of existing/emerging positioning techniques, especially illustrating how CP measurements can lead to centimeter ranging accuracy in indoor factory channels. Giving up the comb structure and cyclic prefix in the positioning reference signal and ensuring its temporal continuity enables noise reduction. Moreover, these directions open new research possibilities for providing phase measurements at the nanosecond-level and introducing high-precision synchronization mechanisms. Yet, using CP in InF scenarios has to overcome challenges by adopting existing solutions from GPS domain related to integer ambiguity and multipath mitigation. The latter is a critical issue where various channels features and fingerprinting techniques can be intelligently incorporated in on-device classification or network-level optimization of positioning schemes.

\bibliographystyle{IEEEtran}
\bibliography{Bibliography.bib}

\begin{thebibliography}{10}
\providecommand{\url}[1]{#1}
\csname url@samestyle\endcsname
\providecommand{\newblock}{\relax}
\providecommand{\bibinfo}[2]{#2}
\providecommand{\BIBentrySTDinterwordspacing}{\spaceskip=0pt\relax}
\providecommand{\BIBentryALTinterwordstretchfactor}{4}
\providecommand{\BIBentryALTinterwordspacing}{\spaceskip=\fontdimen2\font plus
\BIBentryALTinterwordstretchfactor\fontdimen3\font minus \fontdimen4\font\relax}
\providecommand{\BIBforeignlanguage}[2]{{%
\expandafter\ifx\csname l@#1\endcsname\relax
\typeout{** WARNING: IEEEtran.bst: No hyphenation pattern has been}%
\typeout{** loaded for the language `#1'. Using the pattern for}%
\typeout{** the default language instead.}%
\else
\language=\csname l@#1\endcsname
\fi
#2}}
\providecommand{\BIBdecl}{\relax}
\BIBdecl

\bibitem{Dwivedi2021}
S.~Dwivedi \emph{et~al.}, ``Positioning in {5G} networks,'' \emph{IEEE Commun. Mag.}, vol.~59, no.~11, pp. 38--44, 2021.

\bibitem{RP-222616}
{3GPP WID RP-222616}, ``Revised {SID} on study on expanded and improved {NR} positioning,'' Sept. 2022.

\bibitem{38.857}
{3GPP TR 38.857}, ``Study on {NR} positioning enhancements,'' Mar. 2021.

\bibitem{fouda2022toward}
A.~Fouda \emph{et~al.}, ``Toward cm-level accuracy: Carrier phase positioning for {IIoT in 5G}-advanced {NR} networks,'' in \emph{IEEE PIMRC}, 2022.

\bibitem{RD_TOA_Est_TWC}
K.~Shamaei \emph{et~al.}, ``Receiver design and time of arrival estimation for opportunistic localization with {5G} signals,'' \emph{IEEE Trans. Wireless Commun.}, vol.~20, no.~7, pp. 4716--4731, 2021.

\bibitem{Zhang2021}
Z.~Zhang \emph{et~al.}, ``Indoor carrier phase positioning technology based on {OFDM} system,'' \emph{Sensors}, vol.~21, no.~20, 2021.

\bibitem{R1-2104844}
{3GPP Tdoc R1-2104844}, ``Carrier phase based downlink angle of departure measurement,'' May 2021.

\bibitem{Rappaport2021}
O.~Kanhere \emph{et~al.}, ``Position location for futuristic cellular communications: {5G} and beyond,'' \emph{IEEE Commun. Mag.}, vol.~59, no.~1, pp. 70--75, 2021.

\bibitem{Fan2022}
S.~Fan \emph{et~al.}, ``Carrier phase-based synchronization and high-accuracy positioning in {5G} new radio cellular networks,'' \emph{IEEE Trans. Commun.}, vol.~70, no.~1, pp. 564--577, 2022.

\bibitem{mogyorosi2022}
F.~Mogyorósi \emph{et~al.}, ``Positioning in {5G and 6G} networks—{A} survey,'' \emph{Sensors}, vol.~22, no.~13, p. 4757, 2022.

\bibitem{li2023}
X.~Li, ``Gnss repeater based differential indoor positioning with multi-epoch measurements,'' \emph{IEEE Transactions on Intelligent Vehicles}, vol.~8, no.~1, pp. 803--813, 2023.

\bibitem{R1-2104880}
{3GPP Tdoc R1-2104880}, ``Carrier/subcarrier phase based enhancement for {5G NR} positioning,'' May 2021.

\bibitem{Gentile2013}
C.~Gentile \emph{et~al.}, \emph{Multipath and {NLOS} Mitigation Algorithms}, 2013.

\bibitem{R1-2104909}
{3GPP Tdoc R1-2104909}, ``Mitigation of {NLOS} problem for {NR} positioning,'' May 2021.

\bibitem{Sandia2017}
D.~Tso \emph{et~al.}, ``{D-DMTD}: Digital dual mixer time difference,'' {Sandia National Laboratories}, Report SAND2017-10097, August 2017.

\end{thebibliography}

\vspace{-1cm}
\begin{IEEEbiographynophoto}{Jakub Nikonowicz} (jakub.nikonowicz@put.poznan.pl) received the Ph.D. degree in telecommunication systems from Poznań University of Technology (PUT), Poznań, Poland, 2019. Since 2019, where he is currently an assistant professor.
His research interests include statistical signal processing and precise synchronization in distributed systems.
\end{IEEEbiographynophoto}

\vspace{-1cm}
\begin{IEEEbiographynophoto}{Aamir Mahmood} (aamir.mahmood@miun.se) is an assistant professor of communication engineering at Mid Sweden University, Sweden. He received the D.Sc. degrees in communications engineering from Aalto University School of Electrical Engineering, Finland, in 2014. 
His research interests include time synchronization, URLLC, and RAN optimization/management.
\end{IEEEbiographynophoto}

\begin{IEEEbiographynophoto}{Muhammad Ikram Ashraf} (ikram.ashraf@nokia.com) received the M.Sc. and Ph.D. degrees in telecommunication systems and communication engineering, respectively, from the University of Oulu, Finland. He is currently working as a Team lead at Nokia, Finland. His research interests include URLLC, TSN, Positioning, and AI/ML. 
His research interests include URLLC, TSN, Positioning, and AI/ML. 
\end{IEEEbiographynophoto}

\vspace{-1cm}
\begin{IEEEbiographynophoto}{Emil Björnson} (emilbjo@kth.se) received his Ph.D. degree from the KTH Royal Institute of Technology, Stockholm, Sweden, in 2011. He is now a professor of wireless communication at KTH. His research interests include multi-antenna and reconfigurable intelligent surface-aided communications, radio resource allocation, and energy efficiency. 
He is a Fellow of IEEE.
\end{IEEEbiographynophoto}

\vspace{-1cm}
\begin{IEEEbiographynophoto}{Mikael Gidlund} (mikael.gidlund@miun.se) is a professor of computer engineering at Mid Sweden University, Sweden. He has worked as Senior Principal Scientist and Global Research Area Coordinator of Wireless Technologies, ABB Corporate Research, Sweden.
His research interests include wireless communication and networks, access protocols, and security. 
\end{IEEEbiographynophoto}

\end{document}